\documentstyle[graphics]{aa}
\newcommand{\ls}
 {\mathrel{\hbox{\rlap{\hbox{\lower4pt\hbox{$\sim$}}}\hbox{$<$}}}}
\newcommand{\gs}
 {\mathrel{\hbox{\rlap{\hbox{\lower4pt\hbox{$\sim$}}}\hbox{$>$}}}}
\newcommand{\degg}{\hbox{$^\circ$}}

\newcommand{\et}{et al.\ }
\newcommand{\ginga}{{\it Ginga}}

\newcommand{\asca}{{\it ASCA}}

\newcommand{\xmm}{{\it XMM-Newton}}

\def\la{\mathrel{\hbox{\rlap{\hbox{\lower4pt\hbox{$\sim$}}}{\raise2pt\hbox{$
<$}}
}}}
\def\ga{\mathrel{\hbox{\rlap{\hbox{\lower4pt\hbox{$\sim$}}}{\raise2pt\hbox{$
>$}}
}}}

\begin{document}

\thesaurus{(11.01.2; 11.17.4(Markarian 205); 13.25.2)}

\title{XMM-Newton observation of an unusual iron line in the quasar 
Markarian 205}
\author{J.N. Reeves\inst{1} 
\and M.J.L.\ Turner\inst{1}
\and K.A.\ Pounds\inst{1}
\and P.T. O'Brien\inst{1}
\and Th. Boller\inst{2}
\and P. Ferrando\inst{3}
\and E. Kendziorra\inst{4}
\and S. Vercellone\inst{5}}

\offprints{J.N. Reeves}

\institute{X-Ray Astronomy Group; Department of Physics and Astronomy;
Leicester University; Leicester LE1 7RH; U.K.
\and Max-Planck-Institut f{\"u}r extraterrestrische Physik, Postfach 1603,
85748 Garching, Germany
\and Service d'Astrophysique, CEA Saclay, Gif-sur-Yvette, 91191, France
\and Institut f{\"u}r Astronomie und Astrophysik - Astronomie, University
of T{\"u}bingen, Waldh{\"a}user Strasse 64, D-72076 T{\"u}bingen, Germany
\and Istituto di Fisica Cosmica ``G. Occhialini'' - C.N.R., Via Bassini 15,
20133 Milano, Italy}

\date{Received September 2000 / October 2000}

\maketitle

\begin{abstract}

XMM-Newton observations of the low luminosity, radio quiet quasar 
Markarian 205 have revealed a unique iron K emission line
profile. In marked contrast to the broad and redshifted 
iron K line commonly seen in \asca\ observations of Seyfert 1 galaxies,
we find that a substantial amount of the line flux in Mrk 205 occurs above the 
neutral line energy of 6.4~keV. 
Furthermore, we find that the iron line emission has 
two distinct components, a narrow, unresolved neutral
line at 6.4 keV and a broadened line centred at 6.7 keV. 
We suggest that the most 
likely origin of the 6.7 keV line is from 
X-ray reflection off the surface of a highly ionised accretion disk,
whilst the 6.4~keV component may arise from neutral matter distant from the
black hole, quite possibly in the putative molecular torus.
Overall this observation underlines the
potential of \xmm\ for using the iron K line as a diagnostic of matter in
the innermost regions of AGN.


\end{abstract}

\section{Introduction}

Since iron K$\alpha$ emission was found to be a common feature in 
the X-ray spectra of AGN, it has been recognised as potentially a very
important probe of matter in the central nucleus. Observations with
the \ginga\ X-ray satellite first established the
iron K$\alpha$ line (at 6.4 keV) as a dominant feature in the
hard X-ray spectra of many broad-line Seyfert 1 galaxies (Pounds
\et 1990, Nandra \& Pounds 1994). 
The line was often accompanied by a flattening of the
X-ray continuum (the Compton scattering
hump) above the line energy. A widely favoured interpretation of
these spectral features, as primary hard X-rays
being `reflected' by Compton-thick matter - evidence for the
putative accretion disk in AGN - was strengthened when 
\asca, with its higher resolution solid-state detectors,
was able to resolve the iron K emission line. The line profile of the
bright Seyfert 1 galaxy MCG -6-30-15 (Tanaka \et 1995) was found to be 
broadened (v$\sim0.3$c), with a strong asymmetry red-wards of 6.4 keV; the
extreme width and redshift of the iron line being attributed to 
the strong gravity and high velocities of the matter in
the innermost regions around the putative
massive black hole. Subsequent observations of a sample of
Seyfert 1s with \asca\ (Nandra \et 1997) revealed that such
line profiles were apparently common in many nearby Seyfert 1s. 

In contrast, the information on the more luminous, but generally
fainter, quasars has remained inconclusive. 
{\it Ginga} detected iron K emission from 
only a few quasars (Williams \et 1992, Lawson \et
1997), whilst although \asca\ increased the number with
detected iron lines (Reeves \et 1997, Reeves \& Turner 2000), the data 
were not sufficient to resolve any of the line profiles. Despite this,
a divergence in iron K line properties between the low luminosity Seyfert
1 galaxies and higher luminosity quasars has become
apparent, with the line energy being closer to 6.7 keV,
rather than 6.4 keV, and the line strength falling with
increasing luminosity (see Iwasawa \& Taniguichi 1993, Nandra \et
1997b). The sensitivity of the \xmm\ EPIC
instruments now offers the exciting prospect of observing the iron K
emission of quasars in detail and hence studying the flow and physical state of
matter in the nuclei of these most powerful of objects.

This paper presents the discovery, by \xmm, of an unusual  
iron K$\alpha$ line profile in the quasar Markarian 205, 
observed during the Cal-PV (Calibration and Payload
Verification) phase. Mrk 205 is a nearby ($z=0.071$), {\it low
luminosity} quasar (M$_{V}=-23$, L$_{X}=2\times10^{44}$~erg/s,
L$_{BOL}\sim5\times10^{45}$~erg/s; Williams \et 1992; 
Reeves \& Turner 2000) and is radio-quiet (Rush \et 1996). 
It is located only 0.7' away from the nearby ($z=0.00468$)
spiral galaxy NGC 4319 and is viewed through the outer
disk of this galaxy (Bahcall \et 1992, Bowen \& Blades 1993). 
Mrk 205 is also unusual in that it has a very weak UV bump 
(McDowell \et 1989), suggesting that the thermal disc emission is
hidden in the EUV. 
We now present the \xmm\ observations of Mrk 205. 
Values of $H_0=50$~km~s$^{-1}$~Mpc$^{-1}$ and $ q_0 = 0.5 $
are assumed and all fit parameters are given in the quasar rest-frame.

\section{XMM-Newton observations}

Mrk 205 was observed during orbit 75 of the Cal-PV phase of \xmm. 
The observations with the EPIC MOS (Turner \et 2001) and PN
(Str\"{u}der \et 2001) detectors were split
into 3 parts, each of $\sim$17~ksec duration, 
to test a variety of sub-window modes. Three 
observations each were made with the MOS 1 and 2 cameras, in Full Window,
Partial Window 2 and Partial Window 3 modes. For the PN,
two observations were made in Full Window mode and
one in Large Window mode. 
The data were screened with the XMM SAS (Science Analysis 
Software) and pre-processed using the latest CCD gain values known 
at the time of the observation. 
X-ray events corresponding to patterns 0-12 for the 2 MOS cameras 
(similar to grades 0-4 in \asca) were used; for the PN, only 
pattern 0 events (single pixel events) were selected. 
Additional electronic noise was also removed from the MOS detectors.  
A low energy cut of 200 eV was applied 
and known hot or bad pixels were removed during screening. 
The non X-ray background remained low throughout the observations. 
A screened MOS-1 image of Mrk 205 is shown in figure 1.

\begin{figure}
\resizebox{\hsize}{!}{\rotatebox{-90}{\includegraphics{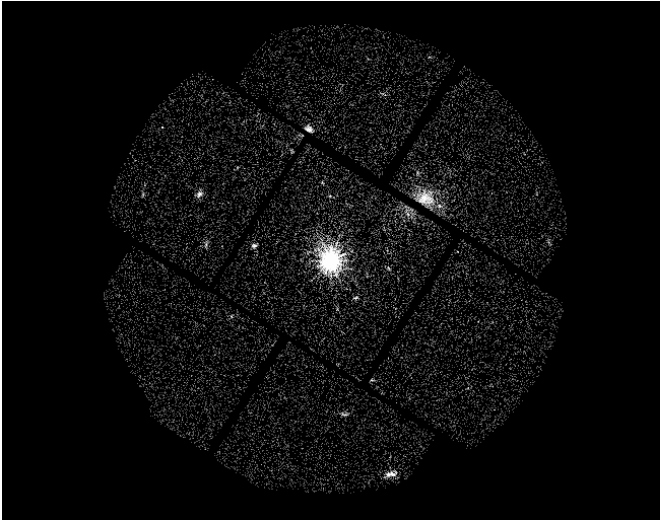}}}
\caption{An EPIC MOS-1 image of the Mrk 205 observation. Mrk 205 is
located in the centre, the extended source near the chip gap is the
nearby elliptical galaxy NGC 4291.}
\end{figure}

We then proceeded to extract spectra and light curves for both the source
and background in each of the three observations, separately for each EPIC
detector. A circular source region of 1' diameter was defined around the
centroid position of Mrk 205, with the background being taken from an
offset position close
to the source. This resulted in 3 source spectra, for each of
the 3 EPIC CCD cameras. There was little flux or spectral variability of
Mrk 205 ($<$10\%) during the observations. Indeed a spectral 
fit of a power-law plus Gaussian emission line (near to 6.4 keV) 
yielded consistent results for each of the three observations. 
Therefore we proceeded to co-add the 3
observations, for each detector. Furthermore as there was little 
difference between the MOS~1 and MOS~2 spectra (both in the spectral
shape and the iron line profile described later), we combined these into a
single spectral file to maximise signal-to-noise. 
The resultant process left one combined EPIC-MOS
spectrum, with a total exposure of 100 ksec (both MOS, co-added) and
one PN spectrum, with a total exposure of 44 ksec. The integrated 
X-ray flux from Mrk 205, over the 0.5-10 keV range, was 
8$\times10^{-12}$~ergs~cm$^{-2}$~s$^{-1}$. At this flux
level, the effect of photon pile-up, even in MOS full-frame mode, is
negligible.
The background subtracted spectra 
were fitted, using \textsc{xspec v11.0}, with the latest response
matrices produced by the EPIC team; the systematic level of uncertainty 
is $<5$\%.  Finally spectra were binned to a minimum of 20 counts per
bin, to apply the $\chi^2$ minimalisation technique.

\section{Spectral Analysis - The iron K line}

We initially fitted the hard X-ray (2.5-10 keV)
spectrum of Mrk 205 with a power law and neutral absorption corresponding
to the line-of-sight 
column through our Galaxy, of $N_{H}=2.8\times10^{20}\rm{cm}^{-2}$
(Elvis \et 1989). 
From our initial inspection of the EPIC spectra, it was
obvious that deviations were present between 6-7 keV suggestive of 
iron K shell band emission. 
We then proceeded to parameterise the profile of the observed feature in
terms of simple 
Gaussian models for both the MOS and PN (see table 1 for a summary of
fits). Note all errors are quoted at the 90\% confidence level. 

\begin{table*}
\centering
\caption{Fits to iron K line profile, in 2.5-10 keV EPIC band.
$^a$ Energy of line in keV. $^b$ Equivalent width of line in eV. $^c$
Intrinsic velocity width of line (in eV) or disk inclination
angle. $^d$ Improvement in $\chi^2$ for adding line
component. $^f$ Indicates parameter value is fixed in the fit. Note PL
= Power-law; GA = Gaussian; DISK = Disk-line. 
For the disk-line, R$_{in}$=6R$_{g}$, R$_{out}$=1000R$_{g}$ and emissivity
$\propto{r}^{-2.5}$ are all assumed.}

\begin{tabular}{@{}llcccccccc@{}}
\hline                 
Fit & Instrument & \multicolumn{3}{c} {Narrow-Line} &
\multicolumn{4}{c} {Broad/Disk Line} & $\chi^{2}$/dof \\   

\ & \ & E$^a$ & EW$^b$ & $\Delta\chi^2$$^d$ & E$^a$ & 
$\sigma$ or $\theta$$^c$ & EW$^b$ &
$\Delta\chi^2$$^d$ \\

\hline

1. PL + 2$\times$GA & MOS only & $6.39\pm0.04$ & $75\pm35$ &
32.4 & $6.68\pm0.22$ & $300^{+280}_{-180}$ & $125^{+95}_{-75}$ & 10.1
& 363.5/341 \\

2. PL + 2$\times$GA & PN only & $6.46^{+0.02}_{-0.08}$ &
$68\pm23$ & 23.7 & $6.78^{+0.12}_{-0.16}$ & $200^{+300}_{-120}$ &
$140^{+110}_{-55}$ & 21.7 & 786/789 \\

3. PL + 2$\times$GA & PN + MOS & $6.39\pm0.03$ & $56\pm23$ & 53 &
$6.74\pm0.12$ & $250^{+190}_{-130}$ & $135^{+70}_{-60}$ & 30 &
1095/1093 \\

4. PL + DISK & PN + MOS & -- & -- & -- & 6.4$^f$ & 45\degg$^f$
& $215^{+80}_{-50}$ & 56 & 1122/1096 \\

5. PL + DISK & PN + MOS & -- & -- & -- & $6.60\pm0.06$ & $42\pm8$\degg
& $300^{+65}_{-70}$ & 68 & 1110/1095 \\

6. PL + GA + DISK & PN + MOS & $6.39\pm0.04$ & $46\pm19$ & 16 &
$6.67\pm0.10$ & $37\pm12$\degg & $225^{+75}_{-105}$ & 68 & 1094/1093 \\

7. PL + 3$\times$GA & PN + MOS & $6.39\pm0.03$ & $70\pm18$ & 50
& $6.59\pm0.06$ & 10$^f$ & $35\pm17$ & 14 \\

\ & \ & \ & \ & \ & $6.85\pm0.05$ & 10$^f$ & $51\pm19$ & 19 &
1095/1092 \\

\hline
\end{tabular}
\end{table*}

Initially we fitted the MOS and PN data separately, to check for
consistency 
between the 2 datasets. For the MOS, the addition 
of a narrow ($\sigma=0.01$~keV) iron K line at $\sim$6.4 keV, to the 
underlying power-law, 
improved the fit significantly ($\Delta\chi^2=32.4$ for 
2 parameters). Even after addition of a narrow iron line, there were still 
residuals present at energies $>6.4$~keV. 
Therefore we added a second line component, with the energy and 
line width both free to vary. The fit improved still further 
($\Delta\chi^2=10.1$ for 3 extra parameters), with the overall 
fit-statistic of $\chi_{\nu}^2=363.5/341$ (fit 1 in table 1). 
The narrow line energy is consistent with neutral iron
(E=$6.39\pm0.04$keV), 
whilst the `broad' Fe line has a higher energy of 6.7$\pm0.2$~keV and an 
rms width of $\sigma=300$~eV (figure 2). 

Next we fitted the PN data, which provided still tighter constraints on
the 
Fe K emission. A significant improvement in the simple power law fit 
($\Delta\chi^2=23.7$ for 2 parameters) was again obtained on adding a 
narrow line near 6.4 keV. 
As for the MOS, the fit was further improved 
($\Delta\chi^2=21.7$ for 3 extra parameters) by adding a second broad
Fe line component (fit 2 in table 1).   
Again the energy of the narrow line is consistent with neutral iron 
(E=6.46$^{+0.02}_{-0.08}$~keV), whereas the energy of the `broad' line 
is near to 6.7 keV (E=$6.78^{+0.12}_{-0.16}$~keV). 
The PN line profile, for the double Gaussian model, is also plotted in
figure 2.  

As the line spectral fits for the MOS and PN are fully consistent, we 
proceeded to fit the 2 spectra simultaneously, allowing for a slight 
($<$5\%) 
normalisation difference between the 2 detectors. The addition of both the 
narrow 6.4 keV line and the broader line at 6.7 keV were both highly 
significant (for the narrow line, $\Delta\chi^2=53$ for 2 parameters; 
broad line, $\Delta\chi^2=30$ for 3 parameters) with the overall 
fit-statistic being very good ($\chi_{\nu}^2=1095.5/1093$, fit 3). 
The narrow line is consistent with emission from neutral iron 
(E=6.39$\pm$0.02~keV, equivalent width EW=56$\pm$23~eV), whilst the broader
line corresponds to ionised matter (E=6.74$\pm0.12$~keV,
$\sigma=250^{+190}_{-130}$~eV and EW=135$^{+65}_{-60}$~eV). 
We do not detect an iron K edge, the upper-limit on the
edge, constrained between 7 and 9 keV, is $\tau<0.2$. However, we
cannot rule out the presence of a continuum reflection component, from
either neutral or moderately ionised material, with a covering
fraction of 0.5 (corresponding to R=$\Omega/2\pi=1$).  
Note the best-fit power-law index of $\Gamma=1.80\pm0.04$ is 
consistent with other radio-quiet quasars (Reeves \et 1997). 
Finally we plot confidence contours between line energy and normalisation 
for both lines in figure 3, illustrating the  
separate detection (at $>$99.9\% confidence) of both the 
narrow line at 6.4 keV and the broad, blue-shifted component 
at 6.7 keV.

\begin{figure}
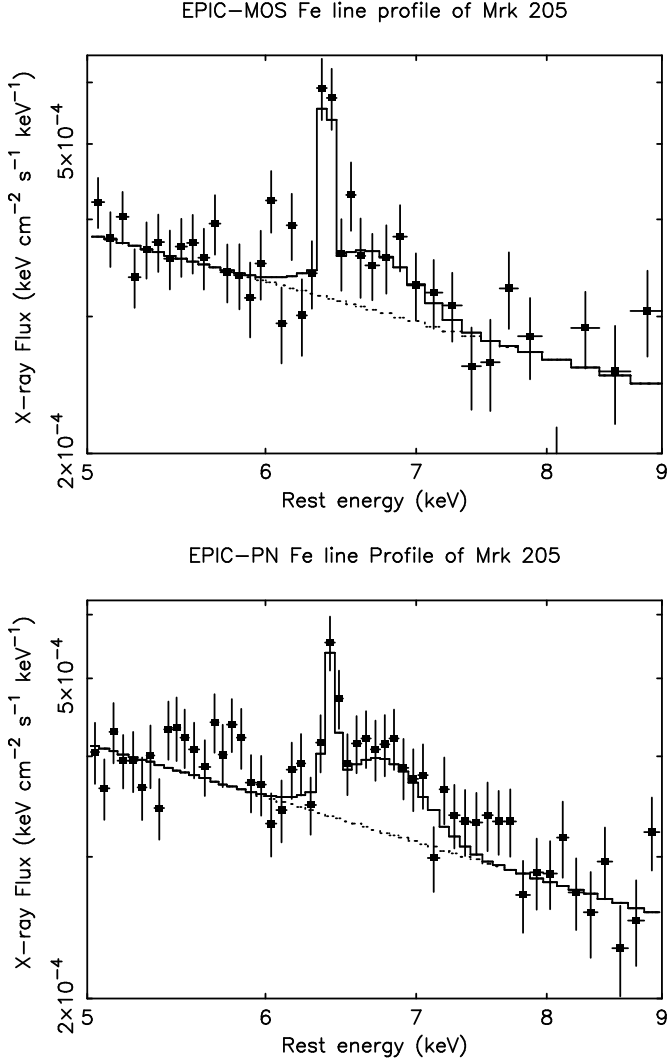

\resizebox{\hsize}{!}{\rotatebox{-90}{\includegraphics{figure2a.ps}}}
\\ 

\resizebox{\hsize}{!}{\rotatebox{-90}{\includegraphics{figure2b.ps}}}
\caption{The EPIC Fe line profile of Mrk 205, from the MOS and PN.  
A broad line component at 6.7 keV is present, 
as well as a narrow emission line at 6.4 keV. The line profile 
differs considerably from those observed in Seyfert 1 galaxies 
(MCG -6-30-15, Tanaka \et 1995; NGC 3516, Nandra \et 1999), 
where most of the line flux is redshifted below 6.4~keV}
\end{figure}

\subsection{Complex fits to the Iron K line Profile}

Given the complexity of the Fe line in Mrk 205, we checked 
whether this could be reconciled with a relativistic disk-line 
profile (Fabian \et 1989). In the first trial, the 
rest-frame energy of the disk-line was fixed at 6.4 keV, appropriate
for neutral iron (i.e. $<$ Fe XX). 
This yielded a poor fit to the
observed  profile (fit 4, table 1), for a disk inclination of 
$\sim$45\degg.  
This is not surprising as the diskline model requires that most of the line
flux is located below 6.4~keV, which is clearly not observed.
Next we freed the rest-frame
energy in the disk-line model (fit 5). This yielded a better fit
($\Delta\chi^{2}=12$), with an increased line energy of E~$=6.60\pm0.06$~keV,
consistent with a higher ionisation of iron. However inspection of the
line residuals indicated significant emission still remaining
at 6.4~keV. Adding a second narrow, neutral iron line to the model
(fit 6) did produce an acceptable fit, with the diskline rest energy
at 6.67$\pm0.10$~keV. {\it Thus the observations are only 
consistent with a diskline origin if the disk ionisation is
high enough to produce He-like iron at $\sim$6.7~keV and  
if the narrow, neutral Fe line component is also present.}

Note that if the 6.7 keV line component does originate from an
ionised accretion disk, then one might still expect to see some line flux
redshifted below 6.4~keV. We therefore tried to add a third, broad emission
line component, with the energy constrained between 5 and 6
keV. However we were only able to place an upper-limit on this
redshifted component (EW~$<45$~eV), and although the flux is
substantially lower it is consistent with the amount
of flux expected from our diskline fit to the ionised line. 
Finally we checked whether a blend of several narrow lines, from various
ionisation species of iron, can model the data (fit 7). This did produce
an acceptable fit $\chi_{\nu}^{2}\sim1$ for three narrow lines, with the
best-fit line energies at 6.4 keV, 6.6 keV and 6.85 keV, compatible
with a line blend from neutral, He-like and H-like iron, respectively.
We now defer further discussion of the iron emission until section 5.

\begin{figure}
\resizebox{\hsize}{!}{\rotatebox{-90}{\includegraphics{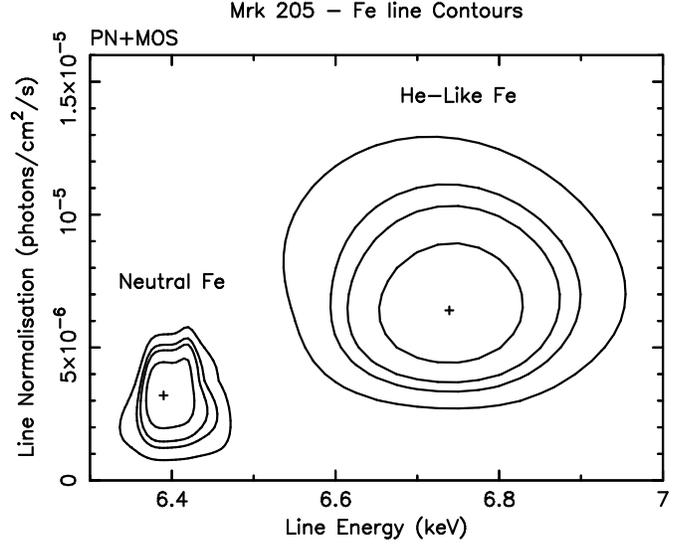}}}
\caption{Confidence contour plot for the double Gaussian Fe line profile. 
Contours correspond to the 68\%, 90\%, 99\% and 99.9\% confidence levels 
respectively. Clearly both the narrow line at 6.4 keV and the broad 
component at 6.7 keV are detected at a high level of significance.}
\end{figure}

\section{The Broad-band XMM spectrum}

\begin{figure}
\resizebox{\hsize}{!}{\rotatebox{-90}{\includegraphics{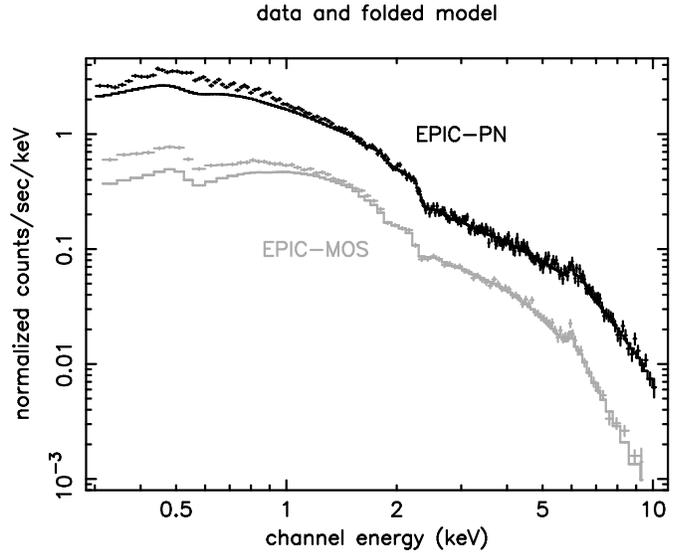}}}
\caption{The EPIC spectrum of Mrk 205, extrapolated to lower energies (0.3
keV). An excess of soft X-ray counts, above that of the hard
power-law, is clearly present in the data.}
\end{figure}

We now consider the broad-band EPIC spectrum of Mrk 205. 
Extrapolating the best-fit hard X-ray spectrum, 
with a power-law index of $\Gamma=1.8$, to lower energies
gives a poor fit; there is a clear excess of soft X-ray counts
above that of the hard X-ray power-law slope (figure 4). 
Refitting the continuum with a single power-law (with
Galactic absorption) still gives a poor fit ($\chi_{\nu}^{2}=1.5$,
1630 channels), albeit with a steeper photon index
($\Gamma=1.96\pm0.01$).  
An acceptable reduced chi-squared ($\chi_{\nu}^{2}=1.2$)
is obtained on adding a black-body component to model the soft
excess, with temperature kT=120$\pm$8 eV and $\Gamma=1.86\pm0.02$. The 
soft excess is relatively weak, representing only $\sim$15\% of the
emission over the range 0.5-2 keV.

Given the apparent weakness of the soft X-ray excess in Mrk 205, and its
relatively high temperature (T=1.5$\times10^{6}$K), it is unlikely that we
are observing the direct thermal emission from the disk in this bandpass
(Malkan \& Sargent 1982). 
A more likely origin of the soft excess in Mrk 205 is by enhanced
reflection of the primary power law continuum in the surface layers of an
ionised accretion disk. Fitting the broad-band
X-ray continuum of Mrk 205 with a hard power-law plus the reflected
contribution from an ionised disk 
(using the \textsc{pexriv} model in XSPEC; Magdziarz \& Zdziarski
1995) provides
a very good fit, including all of the soft X-ray excess. 
The parameters of the fit are also reasonable; R=$\Omega/2\pi=
1.1\pm0.2$,
ionisation parameter $\xi=300\pm80$ erg~cm~s$^{-1}$ and
$\Gamma=2.0\pm0.02$. 
Furthermore this emission could naturally
explain the ionised part (at 6.7 keV) of the iron line profile; He-like
Fe K emission from the disc is predicted in this range of ionisation
(Ross, Fabian \& Young 1999; Nayakshin \et 2000). Note that the detection
of both the narrow and broad Fe line components is still robust (at
$>99.9$\% confidence), even after the addition of reflection to the
continuum. 

Aside from the Fe K line emission, the EPIC X-ray spectrum of Mrk 205
appears
to be devoid of other discrete emission or absorption features. To check
this we
constructed a high resolution RGS spectrum (from RGS1 and RGS2),
using the \textsc{rgsproc} script in the SAS software. 
The RGS spectra are in agreement with the EPIC data, the soft 
X-ray (0.4-2 keV) spectrum of Mrk 205 is
essentially featureless. We place limits on the strength of the OVII
and OVIII edges, expected from a warm absorber, of $\tau<0.12$ and 
$\tau<0.15$ respectively. Furthermore we find
no evidence for neutral photoelectric absorption, other than Galactic; the
limit on
$N_H$ is $<1.5\times10^{20}$~cm$^{-2}$. Thus we are apparently
observing the bare quasar nucleus in Mrk 205. 
There is also no measurable absorption 
from the intervening galaxy NGC 4319, 
consistent with the observation of a very highly ionised ISM 
in the spiral arms of that galaxy (Bowen \& Blades 1993).

\section {The origins of the line in Mrk 205.}

We have found two separate components in the 
Fe K line profile of the quasar Mrk 205, observed with \xmm. 
A narrow line at 6.4 keV is seen, presumably arising 
from neutral material in the quasar rest-frame, as well as 
a broad component at 6.7 keV. 
This observation provides the clearest example to date, from 
CCD-resolution data, of a distinct narrow Fe-K line component in 
a broad-lined, type I AGN. 
We now consider the likely physical origins of the iron line components. 

{\bf The Narrow line at 6.4 keV}. Detection of a strong, narrow iron
emission line at 
6.4 keV implies that a substantial quantity of cool 
reprocessing material is present. This matter 
has to lie outside of the line-of-sight - the X-ray spectrum is 
completely unabsorbed - and therefore is presumably seen by reflection. 
The line strength (EW$\sim50-75$~eV), implies 
that the cool matter subtends a substantial solid angle, 
of at least 1 $\pi$ steradian, assuming it is Thomson thick
and has a solar abundance of iron (e.g. George \& Fabian 1991).  
The most likely location of such material, within the framework of current
AGN models, 
would appear to be the molecular torus (Antonucci 1993), with 
hard X-rays from the central engine being reflected off the 
inner surface of the torus and into the line-of-sight of the observer
(Krolik, Madau \& Zycki 1993).  
Strong iron K$\alpha$ lines are believed to be formed in this way
in Seyfert 2 galaxies 
(Turner \et 1997), where the X-ray continuum 
is at least partly obscured by the putative torus. 
Clear evidence for this torus emission has hitherto not been found in a 
quasar, where the (un-obscured) continuum level is much higher. 
Our observation of a narrow 6.4 keV line from Mrk 205 could provide the
first such evidence. 

Although reflection off the putative torus is attractive, we
cannot rule out other alternatives. Emission from BLR clouds
could 
account for some of the narrow line flux, although the predicted line 
strength is small (EW$<$50~eV; Leahy \& Creighton 1993).  
Interestingly, a high resolution {\it Chandra}-HETG
observation has also resolved a `narrow' Fe K line in the Seyfert 1 NGC
5548 (Yaqoob \et 2000), here the authors attribute the line to BLR
gas, but they cannot exclude the possibility of a torus component. 
Another possibility is that the line originates by X-ray reflection from
the outer 
regions of the accretion disc. Normally the angle subtended by the
outer disk, to the X-ray source, would be small. 
However a warped, concave disk could increase the amount of reflection at
large radii, and account for a measurable narrow line component at
6.4 keV (Blackman 1999). 

{\bf The broad line at 6.7 keV}. The most obvious explanation for the
broad line in Mrk 205 is that 
it originates from the inner accretion disc, where the material
velocities are high, as well as a strong gravity gradient. 
However, the energy of the line emission is {\it not} consistent  
with the diskline profiles observed in many Seyfert 1s 
(Nandra \et 1997), where the line flux is redshifted below 
6.4 keV. Invoking a large inclination angle ($\theta=75-90$\degg) 
can circumvent this, but requires that the disc is orientated
near edge-on, which 
appears to violate current AGN unification ideas (Antonucii 1993). 
An alternative is that the disc matter is in a high 
state of ionisation (with He or even H-like Fe). 
This is a plausible explanation and we note high ionisation Fe lines near
6.7 keV have been claimed from \asca\ data in several radio-quiet quasars 
(Reeves \& Turner 2000), perhaps indicating a level of inner disc
ionisation 
increasing with accretion rate (Matt, Fabian \& Ross 1993). Indeed 
the red-wing of the Fe line profile may disappear if the very 
innermost disc material becomes completely ionised (see Nandra \et
1997b), consistent with our observation of Mrk 205, which we note
is 20$\times$ more luminous than the Seyfert 1 MCG -6-30-15.  

It remains conceivable that the `broad' line does not 
originate from an accretion disc. A spherical rather than planar 
geometry could produce the observed profile, with the blue peak being 
produced naturally through transverse Doppler motion.
Indeed the first formulation of X-ray reflection (Guilbert \& Rees
1988) assumed a spherical distribution of clouds. Again we 
could be observing a blend of several Fe lines, from 
different ionised species and recall this model gave an equally 
good fit to the data (see fit 7).    
One possibility is the ionised lines originate from the 
warm electron scattering region associated with Seyfert 2 galaxies,
which can produce substantial emission from He and H-like iron  
(Krolik \& Kallman 1987). 
We note the spectrum of the Compton-thick Seyfert 2, NGC
1068, shows strong line emission from neutral, He-like and H-like
iron (Marshall \et 1993, Iwasawa \et 1996). 
The difference in Mrk 205, where we observe the direct nuclear X-ray 
continuum, is that the equivalent widths of the lines are 
considerably smaller.

\section{Conclusions}

Observations with \xmm\ have revealed a remarkable iron K line
profile in the low luminosity quasar Mrk 205.  
The broad iron line component at 6.7 keV is
inconsistent with the relativistic profile expected from the inner
accretion disc (Fabian \et 1989), unless the disc material is highly 
ionised. The narrow line at 6.4 keV probably arises from Compton 
scattering off distant cool material;
we may infact be observing the reflected X-ray emission from the putative
molecular torus in Mrk 205.

\section*{ Acknowledgements }
This work is based on observations obtained with XMM-Newton, an ESA science 
mission with instruments and contributions directly funded by 
ESA Member States and the USA (NASA). 
EPIC was developed by the EPIC Consortium led by the Principal 
Investigator, Dr. M. J. L. Turner. The consortium comprises the 
following Institutes: University of Leicester, University of 
Birmingham, (UK); CEA/Saclay, IAS Orsay, CESR Toulouse, (France); 
IAAP Tuebingen, MPE Garching,(Germany); IFC Milan, ITESRE Bologna, 
OAPA Palermo, Italy. EPIC is funded by: PPARC, CEA, CNES, DLR and
ASI.

The authors would also like
to thank the EPIC instrument team, for their hard work during
the calibration phase and the SOC and SSC teams for making the
observation and subsequent analysis possible. It is a pleasure to
thank Gareth Griffiths, Steve Sembay and Richard West for their help and
advice since the launch. We also thank the referee (Paul Nandra) for
providing a rapid reply to this paper and for some useful comments and
suggestions.  


\label{lastpage}

\end{document}